\documentclass[12pt]{article}
\usepackage{amsmath,amssymb}
\usepackage{bm}
\usepackage{graphicx,color}
\begin{document}
\title{
\begin{flushright}
\ \\*[-80pt] 
\begin{minipage}{0.2\linewidth}
\normalsize
\end{minipage}
\end{flushright}
{\Large \bf Searching for   the squark flavor mixing \\
 in CP violations of  $B_s\to K^+ K^-$ and $ \ K^0\bar K^0$ decays
\\*[20pt]}}

\author{Atsushi~Hayakawa$^{1,}$\footnote{E-mail address: hayakawa@muse.sc.niigata-u.ac.jp}, \ \
Yusuke~Shimizu$^{2,}$\footnote{E-mail address:
 yusuke.shimizu@mpi-hd.mpg.de}, \ \ \\
Morimitsu~Tanimoto$^{3,}$\footnote{E-mail address: tanimoto@muse.sc.niigata-u.ac.jp}, \ \  and \ \
Kei~Yamamoto$^{1,}$\footnote{E-mail address: yamamoto@muse.sc.niigata-u.ac.jp}
\\*[20pt]
\centerline{
\begin{minipage}{\linewidth}
\begin{center}
$^1${\it \normalsize
Graduate~School~of~Science~and~Technology,~Niigata~University, \\ 
Niigata~950-2181,~Japan }
\\*[4pt]
$^2${\it \normalsize Max-Planck-Institute f\"ur Kernphysik,
Saupfercheckweg 1, D-69117 Heidelberg, Germany}
\\*[4pt]
$^3${\it \normalsize
Department of Physics, Niigata University,~Niigata 950-2181, Japan }
\end{center}
\end{minipage}}
\\*[70pt]}

\date{
\centerline{\small \bf Abstract}
\begin{minipage}{0.9\linewidth}
\vskip  1 cm
\small
  We study  CP violations in the  $B_s\rightarrow K^+K^-$ and
  $B_s\rightarrow K^0\overline K^0$ decays in order to find the contribution of the supersymmetry,
 which comes from  the gluino-squark mediated  flavor changing current.
 We obtain the allowed region of the squark flavor mixing parameters by putting the experimental data, 
 the mass difference $\Delta M_{B_s}$, the CP violating phase  $\phi_s$ in
 $B_s\to J/\psi  \phi$ decay and  the $b\to s\gamma$ branching ratio.
In addition to these data, we  take into account  the constraint from  the asymmetry of $B^0\rightarrow K^+\pi^-$
because  the $B_s\rightarrow K^+K^-$ decay is related with the  $B^0\rightarrow K^+\pi^-$ decay 
by replacing the spectator $s$ with $d$. 
Under these constraints,  we predict the magnitudes of the  CP violation in  the  $B_s\rightarrow K^+K^-$ and
  $B_s\rightarrow K^0\overline K^0$ decays.
  The predicted region of the CP violation $C_{K^+K^-}$ is strongly  cut
  from the direct CP violation of $\bar{B^0}\to K^-\pi^+$, therefore,
  the deviation from the SM prediction of $C_{K^+K^-}$ is not found. On the other hand, 
 the CP violation $S_{K^+K^-}$ is possibly deviated from the SM prediction considerably, in the region of $0.1\sim 0.5$.
 Since the  standard model  predictions of $C_{K^0 \bar K^0}$ and $S_{K^0 \bar K^0}$
   are very small,  
the squark contribution can be detectable 
in $C_{K^0 \bar K^0}$ and $S_{K^0 \bar K^0}$.  These  magnitudes are expected 
 in the region  $C_{K^0 \bar K^0}=-0.06\sim 0.06$ and $S_{K^0 \bar K^0}=-0.5\sim 0.3$.
 More precise data of these CP violations  provide us a crucial test for the gluino-squark mediated  flavor changing current.
\end{minipage}
}

\begin{titlepage}
\maketitle
\thispagestyle{empty}
\end{titlepage}

\section{Introduction}
\label{sec:Intro}

Recently, there have been a lot of studies to search for new physics 
in the low energy flavor physics such as $B_s$ decays.
Actually, the LHCb collaboration has reported 
new data of  the CP violations of the $B_s$ meson and the branching ratios 
of rare $B_s$ decays~\cite{Bediaga:2012py}-\cite{LHCb:2011ab}.
For many years, the CP violations in the $K$ and $B^0$ mesons 
have been successfully understood within the framework of the standard model (SM), 
so called Kobayashi-Maskawa (KM) model \cite{Kobayashi:1973fv}, 
where the source of the CP violation is the KM phase in the quark sector with three families. 
However, the new physics has been expected to be indirectly discovered
in the flavor changing neutral current (FCNC) of the  $B^0$ and $B_s$ decays
 at the LHCb experiment and the further coming  experiment Belle II.

The LHCb collaboration presented the  data of  the time dependent CP asymmetry 
in the non-leptonic $B_s\to {J/\psi \phi}$ decay \cite{Aaij:2013oba,LHCb:2011aa,LHCb:2011ab}, which is consistent with the SM prediction.  Therefore, this observed value gives us a strong constraint of the new physics contribution to the $b\to s$ transition. 
 In addition to this result,  the first measurement of time-dependent
CP violation in $B_s\rightarrow K^+K^-$ decay has been reported at LHCb \cite{Aaij:2013tna}.
Some authors discussed this process and  the $B_s\rightarrow K^0 \overline K^0$ one in order to search for new physics 
 \cite{Fleischer:2004vu}-\cite{Bhattacharya:2012hh},
because the penguin amplitudes dominate these decays.
Especially,  the  SM prediction of the CP violation of the  $B_s\rightarrow K^0 \overline K^0$ decay
   is  very small,  and so,  the new physics contribution can be detectable in the time dependent CP asymmetry.
   
On the other hands, it is noticed that the $B_s\rightarrow K^+K^-$ decay is related with 
 the  $B^0\rightarrow K^+\pi^-$ decay
by replacing the spectator $s$ with $d$.
Thus, the  $B^0\rightarrow K^+\pi^-$ decay associates with 
the processes of $B_s\rightarrow K^+K^-$ and  $B_s\rightarrow K^0 \overline K^0$
in order  to search for the new physics in the $b\to s$ penguin process.
 It is found that the recent experimental data of the direct CP violation in $B^0\rightarrow K^+\pi^-$ decay 
 is well agreement with the SM prediction with the QCD factorization calculation
\cite{Beneke:1999br,Beneke:2000ry}.
This process depends on the form factor $F(B\rightarrow K)$ and the chiral enhancement
factor $(2M_K^2/m_b m_s)$ in the framework of the QCD factorization.
 The amplitudes of $B_s\rightarrow K^+K^-$ and
  $B_s\rightarrow K^0\overline K^0$ decays also
  involve the common  form factor and chiral enhancement
 factor under  neglecting   the difference of masses of the  $B^0$ and $B_s$ mesons.

 As the new physics, we examine the sensitivity of the effect of the supersymmetry (SUSY)
  in the  CP violation of these  $B_s$ decays.
Although the SUSY is one of the most attractive candidates for the new physics,
the SUSY signals have not been observed yet.  
Since the lower bounds of the superparticle masses increase gradually, 
the squark and the gluino masses are supposed  
to be at the TeV scale~\cite{squarkmass}. 
While, there are new sources of the CP violation in the low energy flavor physics 
if the SM is extended to the SUSY model. The soft squark mass matrices contain 
the CP-violating phases, which contribute to the FCNC with the CP violation. 
Therefore, one expects the effect of the SUSY contribution in the CP-violating phenomena of the $B_s$ meson decays. 
We study the gluino-squark mediated  flavor changing process, which is  the most important process of the SUSY contribution for the  $b\to s$ transition \cite{King:2010np}- \cite{Shimizu:2013jia}.   
 
The gluino mass is  expected 
to be larger than $1.3$~TeV, and the squarks of the first and second
 families are also heavier than $1.4$~TeV~\cite{squarkmass}. 
 Therefore, we take the split-family scenario, in which the first and second family
 squarks are very heavy, ${\cal O}(10-100)$~TeV, while the third family
  squark masses are at  ${\cal O}(1)$~TeV.
  Then,  the  $s\to d$ transition mediated by the first and second family squarks
  is  suppressed by their heavy masses, and competing process is mediated by the second order contribution of the third family squark.
  In order to estimate the gluino-squark mediated FCNC 
  for the  $B_s$ meson decays,
  we work in the basis of  the squark mass eigenstate. 
Then, the $6\times 6$ mixing matrix among down-squarks and down-quarks is discussed
 by input of the experimental constraints. 

In section 2, we present the formulation of the CP violation  of
the $B^0$ and $B_s$ decays in the QCD factorization. In section 3, we present the setup
in our  split-family scenario.
In section 4, we discuss the sensitivity of the gluino-squark mediated FCNC to the  CP violation of 
 the $B^0\rightarrow K^+\pi^-$, $B_s\rightarrow K^+K^-$ and, $B_s\rightarrow K^0 \overline K^0$ decays.
Section 5 is devoted to the summary. 
Relevant formulations are presented in appendices A, B, and C.

\section{ CP violation of $B$ decays in QCD factorization}
In this section, we present the formulation of the CP violation in 
 $B^0\rightarrow K^+\pi^-$, $B_s\rightarrow K^+K^-$, and $B_s\rightarrow K^0 \overline K^0$
 decays within the framework of the QCD factorization~
\cite{Beneke:1999br,Beneke:2000ry,Muta:2000ti,Giri:2004af}. 
First, we begin with the effective Hamiltonian for the $\Delta B=1$ transition as
\begin{equation}
H_{eff}=\frac{4G_F}{\sqrt{2}}\left [\sum _{q'=u,c}V_{q'b}V_{q'q}^*
\sum _{i=1,2}C_iO_i^{(q')}-V_{tb}V_{tq}^*
\sum _{i=3-10,7\gamma ,8G}\left (C_iO_i+\widetilde C_i\widetilde O_i\right )\right ],
\label{hamiltonian}
\end{equation}
where $q=s,d$. The local operators are given as 
\begin{align}
&O_1^{(q')}=(\bar q_\alpha\gamma _\mu P_Lq_\beta')
(\bar q_\beta'\gamma ^\mu P_Lb_\alpha),
\qquad O_2^{(q')}=(\bar q_\alpha\gamma _\mu P_Lq_\alpha')
(\bar q_\beta'\gamma ^\mu P_Lb_\beta), \nonumber \\
&O_3=(\bar q_\alpha\gamma _\mu P_Lb_\alpha)\sum _Q(\bar Q_\beta\gamma ^\mu P_LQ_\beta),
\quad O_4=(\bar q_\alpha\gamma _\mu P_Lb_\beta)\sum _Q(\bar Q_\beta\gamma ^\mu P_LQ_\alpha), \nonumber \\
&O_5=(\bar q_\alpha\gamma _\mu P_Lb_\alpha)\sum _Q(\bar Q_\beta\gamma ^\mu P_RQ_\beta),
\quad O_6=(\bar q_\alpha\gamma _\mu P_Lb_\beta)\sum _Q(\bar Q_\beta\gamma ^\mu P_RQ_\alpha), \nonumber \\
&O_7=\frac{3}{2}(\bar q_\alpha\gamma _\mu P_Lb_\alpha)\sum _Q(e_Q\bar Q_\beta\gamma ^\mu P_RQ_\beta),
\quad O_8=\frac{3}{2}(\bar q_\alpha\gamma _\mu P_Lb_\beta)\sum _Q(e_Q\bar Q_\beta\gamma ^\mu P_RQ_\alpha), \nonumber \\
&O_9=\frac{3}{2}(\bar q_\alpha\gamma _\mu P_Lb_\alpha)\sum _Q(e_Q\bar Q_\beta\gamma ^\mu P_LQ_\beta),
\quad O_{10}=\frac{3}{2}(\bar q_\alpha\gamma _\mu P_Lb_\beta)\sum _Q(e_Q\bar Q_\beta\gamma ^\mu P_LQ_\alpha), \nonumber \\
&O_{7\gamma }=\frac{e}{16\pi ^2}m_b\bar q_\alpha\sigma ^{\mu \nu }P_Rb_\alpha
F_{\mu \nu }, 
\qquad O_{8G}=\frac{g_s}{16\pi ^2}m_b\bar q_\alpha\sigma ^{\mu \nu }
P_RT_{\alpha\beta}^ab_\beta G_{\mu \nu }^a,
\end{align}
where $P_R=(1+\gamma _5)/2$, $P_L=(1-\gamma _5)/2$, and $\alpha $, $\beta $ are color 
indices, and $Q$ is taken to be $u,d,s,c$ quarks. 
Here, $C_i$'s and $\widetilde C_i$'s are the Wilson coefficients at the relevant mass scale, 
and $\widetilde O_i$'s are the operators by replacing $L(R)$ with $R(L)$ 
in $O_i$. The $\widetilde C_i$'s  are neglected in SM. 

 We use the value of Wilson coefficients at $\mu=m_b$ as follows:
\begin{align}
 &C_1=-0.185, \quad C_2=1.082, \quad C_3=0.014, \quad
 C_4=-0.035,  \nonumber\\
& C_5=0.009, \quad C_6=-0.041, \quad  C_7=-0.002/137, \quad C_8=0.054/137, \nonumber \\  &C_9=-1.292/137, \quad
 C_{10}=-0.262/137, \quad C_{8G}=-0.143,
\end{align}
in the SM calculations \cite{Muta:2000ti}.

The hard scattering amplitude is given for the relevant decay modes as follows:
\begin{align}
\mathcal{T}_p=
&\frac{4G_F}{\sqrt{2}} \sum_{p=u,c} V_{pq}^{*}V_{pb} 
 \Big[ a_1^p (\bar q \gamma_{\mu} L u)\otimes(\bar u \gamma^{\mu} L b)
 +a_2^p (\bar u \gamma_{\mu} L u)\otimes(\bar q \gamma^{\mu} L b)  
 +a_3^p (\bar {q'} \gamma_{\mu} L q')\otimes(\bar q \gamma^{\mu} L b)  \nonumber\\
&+a_4^p (\bar {q} \gamma_{\mu} L q')\otimes(\bar q \gamma^{\mu} L b)
 +a_5^p (\bar {q'} \gamma_{\mu} R q')\otimes(\bar q \gamma^{\mu} L b)
 +a_6^p (-2)(\bar {q} R q')\otimes(\bar{q'}L b) \nonumber\\
&+a_7^p \frac{3}{2} e_{q'}(\bar{q'} \gamma_{\mu}R q')\otimes(\bar{q} \gamma^{\mu}L b)
 +(-2)(a_8^p \frac{3}{2} e_{q'}+a_{8a})(\bar{q}R q')\otimes(\bar{q'} L b) \nonumber\\
 &+a_9^p \frac{3}{2} e_{q'}(\bar{q'}\gamma_{\mu}L q')\otimes(\bar{q}\gamma^{\mu} L b) 
+(a_{10}^p \frac{3}{2} e_{q'}+a_{10a}^p)(\bar{q}\gamma_{\mu}L q')\otimes(\bar{q'}\gamma^{\mu} L b) \Big],
\end{align}
where the symbol $\otimes$ denotes $\langle M_1M_2|j_2\otimes j_1|B\rangle \equiv \langle M_2|j_2|0\rangle \langle M_1|j_1|B\rangle$.
The effective $a_i^p$'s which contain next-to leading order (NLO) coefficients and
${\cal O}(\alpha_s)$ hard scattering corrections are given as,
\begin{align}
a_{1,2}^c &= 0,  \qquad
a_i^c=a_i^u \ (i = 3,5,7,8,9,10,8a,10a),  \qquad
a_1^u 
= C_2 +\frac{C_1}{N}+\frac{\alpha_s}{4\pi}\frac{C_F}{N}C_1 F_{M_2}, \nonumber  \\
a_2^u 
&= C_1 +\frac{C_2}{N}+\frac{\alpha_s}{4\pi}\frac{C_F}{N}C_2 F_{M_2},  \qquad
a_3^u 
= C_3 +\frac{C_4}{N}+\frac{\alpha_s}{4\pi}\frac{C_F}{N}C_4 F_{M_2},  \nonumber \\
a_4^p 
&= C_4 +\frac{C_3}{N}+\frac{\alpha_s}{4\pi}\frac{C_F}{N}\Big[ C_3 \big[ F_{M_2}+G_{M_2}(s_q)+G_{M_2}(s_b) \big] + C_2 G_{M_2}(s_q) \nonumber \\
&+(C_4+C_6)\sum_{f=u}^b G_{M_2}(s_f)+C_{8_G}G_{M_{2},g} \Big],  \nonumber \\
a_5^u 
&= C_5 +\frac{C_6}{N}+\frac{\alpha_s}{4\pi}\frac{C_F}{N}C_6(-F_{M_2}-12),  \nonumber \\
a_6^p 
&= C_6 +\frac{C_5}{N}+\frac{\alpha_s}{4\pi}\frac{C_F}{N}\Big[ C_2 G_{M_2}'(s_p) +C_3\big[ G_{M_2}'(s_q)+G_{M_2}'(s_b) \big] \nonumber \\
& +(C_4+C_6)\sum_{f=u}^b G_{M_{2}}'(s_f)+C_{8G}G_{M_{2},g}' \Big],  \nonumber \\
a_7^u 
&= C_7 +\frac{C_8}{N}+\frac{\alpha_s}{4\pi}\frac{C_F}{N}C_8(-F_{M_2}-12),   \qquad
a_8^p 
= C_8 +\frac{C_7}{N},  \nonumber \\
a_{8a}^p 
&= \frac{\alpha_s}{4\pi}\frac{C_F}{N}\Big[ (C_8+C_{10})\sum_{f=u}^{b} \frac{3}{2}e_fG_{M_2}'(s_f)+C_9\frac{3}{2}[e_q G_{M_2}'(s_q)+e_bG_{M_2}'(s_b)] \Big],\nonumber \\
a_9^u 
&= C_9 +\frac{C_{10}}{N}+\frac{\alpha_s}{4\pi}\frac{C_F}{N}C_{10}F_{M_2},  \qquad
a_{10}^u 
= C_{10} +\frac{C_9}{N}+\frac{\alpha_s}{4\pi}\frac{C_F}{N}C_9F_{M_2},  \nonumber  \\
a_{10a}^p 
&= \frac{\alpha_s}{4\pi}\frac{C_F}{N}\Big[ (C_8+C_{10})\sum_{f=u}^{b} \frac{3}{2}e_fG_{M_2}(s_f)+C_9\frac{3}{2}[e_q G_{M_2}(s_q)+e_bG_{M_2}(s_b)] \Big],
\label{ai}
\end{align}
where $q=d,s \ \ q'=u,d,s \ \ f=u,d,s,c,b$ and $C_{F}=(N^2-1)/(2N)$ with  the number of colors $N=3$.
In Appendix A, we present the loop integral functions  $F_{M_2}$ $G_{M_{2,g}}$, $G_{M_2}(s_q)$, $G_{M_{2,g}}'$
and $G_{M_2}'(s_q)$, in which 
the internal quark mass enters as $s_f=m_f^2/m_b^2$.

In this work, $C_i$ includes both SM contribution and squark-gluino one, 
such as $C_i=C_i^{\rm SM}+C_i^{\tilde g}$, where 
$C_i^{\text{SM}}$'s are given in Ref.~\cite{Buchalla:1995vs}. 
The Wilson coefficients of the gluino-squark contribution 
$C_{7\gamma}^{\tilde g}$ and $C_{8G}^{\tilde g}$ are presented in Appendix B. 
We should also take account of the SUSY contribution 
in   $\widetilde C_i$'s$(i=3-10,7_\gamma, 8_G)$,  which are 
derived  by replacing $L(R)$ with $R(L)$ in $C_i$. 
Then, $C_i$'s are replaced with  $C_i-\widetilde C_i$ in Eq.(\ref{ai})
for the decays $B_s\rightarrow K^+K^-$ and $B_s\rightarrow K^0 \overline K^0$. 
The minus sign in front of $\widetilde C_i$ is due to the parity of the final states.

By using these formula, we can write the decay amplitude  
for the $\bar{B^0}\to K^-\pi^+$, $\bar{B_s}\to K^+K^-$ and $\bar{B_s}\to K^0\overline K^0$ decays, 
respectively, as follows:
\begin{align} 
\mathcal{\bar A}&(\bar{B^0}\to K^-\pi^+)
=
\frac{G_F}{\sqrt 2}if_{\pi}(M_{B^0}^2-M_K^2)F^{B^0\to K}(0)(1-\frac{\lambda^2}{2})|V_{cb}|\Big(R_{CKM} e^{-i\gamma}\big[ a_1^u+a_4^u \nonumber \\ 
&+R_K (a_6^u+a_8^u+a_{8a}) +a_{10}^u+a_{10a}^u \big]+ \big[a_4^c+R_K(a_6^c+a_8^c)+a_{10}^c+a_{10a}^c] \Big),
\end{align} 
\begin{align} 
\mathcal{\bar A}&(\bar{B_s}\to K^+K^-)
=
\frac{G_F}{\sqrt 2}if_{K}(M_{B_s}^2-M_K^2)F^{B_s\to K}(0)(1-\frac{\lambda^2}{2})|V_{cb}|\Big( R_{CKM} e^{-i\gamma}\big[ a_1^u+a_4^u \nonumber \\
 &+R_K (a_6^u+a_8^u+a_{8a}) +a_{10}^u+a_{10a}^u \big]+ \big[a_4^c+R_K(a_6^c+a_8^c)+a_{10}^c+a_{10a}^c] \Big),
\end{align} 
\begin{align} 
\mathcal{\bar A}&(\bar{B_s}\to K^0 \bar K^0)
=
\frac{G_F}{\sqrt 2}if_{K}(M_{B_s}^2-M_K^2)F^{B_s\to K}(0)(1-\frac{\lambda^2}{2})|V_{cb}|\Big( R_{CKM} e^{-i\gamma}\big[a_4^u \nonumber\\
 &+R_K (a_6^u+a_8^u+a_{8a}) +a_{10}^u+a_{10a}^u \big]+ \big[a_4^c+R_K(a_6^c+a_8^c)+a_{10}^c+a_{10a}^c] \Big),
\end{align} 
where
\begin{align*} 
R_{CKM} = \frac{\lambda}{1-\lambda^2/2}\left| \frac{V_{ub}}{V_{cb}} \right|,
\end{align*} 
and  $f_{\pi(K)}$,  $F^{B^0(B_s)\to K}(0)$ are decay constants and the form factors at $q^2=0$, respectively.
The CKM matrix elements  $V_{cb},V_{ud}$ and $V_{us}$ are chosen to be real and $\gamma$ is the phase of $V_{ub}^*$, and 
we take $\lambda=V_{us}=0.22535$ and  $R_K=2M_K^2 / ((m_s+m_{\bar d})(m_b-m_q))$.

Let us discuss the time dependent CP asymmetries of $B_s$ decaying 
into the final state $f$, which are defined as~\cite{Aushev:2010bq} 
\begin{equation}
C_f= \frac{1-|\lambda_{f}|^2}{1+|\lambda_{f}|^2}\ , 
\qquad S_f=\frac{2\text{Im}\lambda _{f}}{1+|\lambda_{f}|^2}\ ,
\label{sf}
\end{equation}
where 
\begin{equation}
\lambda_{f}=\frac{q}{p} \bar \rho\ , \qquad 
\frac{q}{p}\simeq \sqrt{\frac{M_{12}^{s*}}{M_{12}^{s}}}, \qquad 
\bar \rho \equiv 
\frac{\bar A(\bar B_s\to f)}{A(B_s\to f)}.
\label{lambdaf}
\end{equation}

In the  $B_s\to J/\psi  \phi$ decay,
 we write  $\lambda_{J/\psi  \phi}$ in terms of phase factors as follow:
\begin{equation}
\lambda _{J/\psi \phi } \equiv e^{-i\phi _s}.
\label{new}
\end{equation}
In the SM, the angle $\phi_s$ is given as $\phi_s=-2\beta_s$,
in which $\beta_s$ is one angle  of  the unitarity triangle   for $B_s$.
The SM predicts $\phi_s $ as \cite{Charles:2004jd}
\begin{equation}
\phi_s=-0.0363\pm 0.0017\ .
\end{equation}

The recent experimental data of this phase is \cite{Aaij:2013oba,Amhis:2012bh}
\begin{equation}
 \phi_s=0.07\pm 0.09\pm 0.01 \ .
\label{phasedata}
\end{equation}
 This value  constrains  the magnitude of the new physics,
 which  contributes to $M_{12}^s$ in Eq.(\ref{lambdaf}).
 For  the gluino-squark  contribution to $M_{12}^s$, we present the formulation in Appendix C.

The time dependent CP asymmetries of $B_s\to K^+K^-$ and  $B_s\to K^0 \bar K^0$
are obtained by calculating
\begin{equation}
\lambda _{K^+ K^-}=e^{-i\phi_s}\ 
\frac{\mathcal{\bar A}(\bar B_s \to K^+ K^-)}{\mathcal{A}(B_s\to K^+  K^-)}\ , \qquad
\lambda _{K^0\bar K^0}=e^{-i\phi_s}\ 
\frac{\mathcal{\bar A}(\bar B_s \to K^0 \bar K^0)}{\mathcal{A}(B_s\to K^0 \bar K^0)}\ .
\end{equation}


The new physics contribution is often  sensitive in the $b\to s\gamma $ decay. 
The branching ratio BR$(b\to s\gamma )$ is given as~\cite{Buras:1998raa} 
\begin{equation}
\frac{\text{BR}(b\to s\gamma )}
{\text{BR}(b\to ce\bar {\nu _e})}
=
\frac{|V_{ts}^*V_{tb}|^2}
{|V_{cb}|^2}
\frac{6 \alpha }{\pi f(z)}
(|C_{7\gamma }(m_b)|^2+|{\tilde C}_{7\gamma }(m_b)|^2),
\label{Brbqgamma}
\end{equation}
where 
\begin{equation}
f(z)
=
1-8z+8z^3-z^4-12z^2 \text{ln}z~,\qquad 
z = \frac{m_{c,pole}^2}{m_{b,pole}^2}.
\end{equation}
Here $C_{7\gamma }(m_b)$ and $\tilde{C}_{7\gamma }(m_b)$ include both contributions 
from the SM and the new physics. 
The SM prediction including the next-to-next-to-leading order correction is given as \cite{Misiak:2006zs}
\begin{equation}
\text{BR}(b\to s\gamma )({\rm SM})=(3.15 \pm 0.23)\times 10^{-4},
\end{equation}
on the other hand, the experimental data is obtained as \cite{PDG} 
\begin{equation}
\text{BR}(b\to s\gamma )({\rm exp})=(3.53 \pm 0.24)\times 10^{-4}. 
\end{equation}
By inputing this experimental value, the contribution of the gluino-squark mediated flavor changing process,
$C_{7\gamma}$ and $\tilde C_{7\gamma}$,  is constrained.

  In addition to the CP violating processes with  $\Delta B=2,\ 1$, the SUSY contribution  is also sensitive to
the electric dipole moment \cite{Fuyuto:2013gla}, which is the the T violation of the flavor conserving process.
The experimental upper bound of the  electric dipole moment of the neutron
provides us the upper-bound of  the chromo-EDM(cEDM) of the strange quark 
\cite{Hisano:2003iw}-\cite{Fuyuto:2012yf}. 
The cEDM of the strange quark $d_{s}^C$ is given 
in terms of the gluino-sbottom-quark interactions \cite{Shimizu:2013jia}.
The upper bound of the cEDM of the strange quark is given
by the experimental upper bound of the neutron EDM as \cite{Fuyuto:2012yf},
\begin{equation}
 e|d_s^C|<0.5\times 10^{-25} \ \text{ecm}.
 \label{cedm}
 \end{equation}
 
 This bound  constrains  the SUSY flavor mixing angles and the phases in $C_{8G}$ and $\tilde C_{8G}$.
However, the experimental data of the direct CP violation in the $B^0\rightarrow K^+\pi^-$ decay
gives a little bit stronger constraint for  $C_{8G}$ and $\tilde C_{8G}$ in our framework.
Therefore, we omit the discussion about the cEDM in this work.


\section{ Setup of squark flavor mixing}
\label{sec:Deviation}
Let us discuss the gluino-squark mediated flavor changing process 
as the dominant SUSY contribution of the $b\to s$ transition. We give the $6\times 6$ squark mass matrix to be $M_{\tilde q}$
$(\tilde q=\tilde u, \tilde d)$  in the super-CKM basis. In order to go to the  diagonal basis of the squark mass matrix,
 we rotate $M_{\tilde q}$ as
\begin{equation}
 \tilde m_{\tilde q, \rm diagonal}^2=\Gamma _{G}^{(q)}M_{\tilde q}^2 \Gamma _{G}^{(q)  \dagger} \ ,
\end{equation}
where  $\Gamma _{G}^{(q)}$ is the $6\times 6$ unitary matrix, and
we decompose it into the  $3\times 6$ matrices 
 as $\Gamma _{G}^{(q)}=(\Gamma _{GL}^{(q)}, \ \Gamma _{GR}^{(q))})^T$ in the following expressions. 
Then, the gluino-squark-quark interaction is given as
 \begin{equation}
\mathcal{L}_\text{int}(\tilde gq\tilde q)=-i\sqrt{2}g_s\sum _{\{ q\} }\widetilde q_i^*(T^a)
\overline{\widetilde{G}^a}\left [(\Gamma _{GL}^{(q)})_{ij}{\bm L}
+(\Gamma _{GR}^{(q)})_{ij}{\bm R}\right ]q_j+\text{h.c.}~,
\end{equation}
where $\widetilde G^a$ denotes the gluino field, and  ${\bm L}$ and ${\bm R}$ are projection operators. 
This interaction leads to the gluino-squark mediated flavor changing process 
with $\Delta B=2$ and  $\Delta B=1$ 
through the  box and  penguin diagrams.

We take the split-family scenario, in which the first and second family
 squarks are very heavy, ${\cal O}(10-100)$~TeV, while the third family
  squark masses are at  ${\cal O}(1)$~TeV.
Therefore, the first and second squark contribution is  suppressed
in the gluino-squark mediated flavor changing process by their heavy masses. 
In addition, we also assume the flavor symmetry such as U(2) \cite{U2}
 in order to suppress FCNC enough in the neutral K meson system \cite{Gabbiani:1996hi}.  
 The stop and sbottom interactions dominate the gluino-squark mediated flavor changing process. 
Then, the sbottom interaction contributes  $\Delta B=2$ and  $\Delta B=1$ processes.
 We take a suitable parametrizations of
 $\Gamma _{GL}^{(d)}$ and $\Gamma _{GR}^{(d)}$ as follows \cite{Mescia:2012fg}:
 \begin{align}
\Gamma _{GL}^{(d)}&=
\begin{pmatrix}
1 & 0 & \delta _{13}^{dL}c_\theta & 0 & 0 & -\delta _{13}^{dL}s_\theta e^{i\phi } \\
0 & 1 & \delta _{23}^{dL}c_\theta & 0 & 0 & -\delta _{23}^{dL}s_\theta e^{i\phi } \\
-{\delta _{13}^{dL}}^* & -{\delta _{23}^{dL}}^* & c_\theta & 0 & 0 & -s_\theta e^{i\phi }
\end{pmatrix}, \nonumber \\
\nonumber\\
\Gamma _{GR}^{(d)}&=
\begin{pmatrix}
0 & 0 & \delta _{13}^{dR}s_\theta e^{-i\phi } & 1 & 0 & \delta _{13}^{dR}c_\theta \\
0 & 0 & \delta _{23}^{dR}s_\theta e^{-i\phi } & 0 & 1 & \delta _{23}^{dR}c_\theta \\
0 & 0 & s_\theta e^{-i\phi } & -{\delta _{13}^{dR}}^* & -{\delta _{23}^{dR}}^* & c_\theta 
\end{pmatrix},
\label{mixing}
\end{align}
where $c_\theta =\cos \theta$ and $s_\theta =\sin \theta$, 
with the mixing angle  $\theta$ in the $\tilde b_{L,R}$ sector 
and $\delta _{j3}^{dL}$, $\delta _{j3}^{dR}$ are the couplings responsible for the flavor transitions. 
The mixing angle $\theta$ comes from the trilinear SUSY breaking terms. 
If this breaking is neglected,  $\theta$ vanishes. 
In our work, we suppose the large $\mu\tan\beta$, which leads to the non-negligible  mixing angle $\theta$ in the $\tilde b_L- \tilde b_R$ sector.
By using these rotation matrices, we estimate the gluino-sbottom mediated flavor changing amplitudes in the  $B_s$ meson decay.

For the numerical analysis, we fix  sbottom masses.
The third family squarks can have substantial mixing between the left-handed squark and the right-handed one due to 
the large Yukawa coupling, that is the large   $\mu\tan\beta$.
 In our numerical calculation, we take the typical mass eigenvalues $m_{\tilde b_1}$  and  $m_{\tilde b_2}$, and the gluino mass $m_{\tilde g}$ as follows:
  \begin{equation}
  m_{\tilde b_1}=1 \ {\rm TeV}, \qquad m_{\tilde b_2}=1.5 \ {\rm TeV}, 
  \qquad m_{\tilde g}=2 \ {\rm TeV},
  \label{setup}
 \end{equation}
  where we take account of the present experimental bounds \cite{squarkmass}.
  Once we fix mass eigenvalues $m_1$, $m_2$ and $\mu\tan\beta$, 
  we can  estimate the mixing angle $\theta$  between the left-handed sbottom and the right-handed one \cite{Martin:1997ns}. 
Taking  $\mu\tan\beta=20-50$ TeV,  we estimate  $\theta$ in the range of $4^\circ-10^\circ$, which is used
 in our numerical calculations.
  If we take $\mu\tan\beta\ll 20$ TeV, the left-right mixing angle $\theta$ is much less than ${\cal O}(1^\circ)$. Then, 
 the SUSY contribution in $C_{8G}$ and  $C_{7\gamma}$ are tiny because  the left-right mixing dominates $C_{8G}$ and  $C_{7\gamma}$. 
 The smaller mass difference  $m_{\tilde b_2}- m_{\tilde b_1}$ gives  the larger mixing  angle $\theta$.
 However,  our results does not so change since the SUSY contribution
 depends on the combination of $\theta $ and the mass deference as 
$\sin 2\theta \times (m_{\tilde b_2}^2- m_{\tilde b_1}^2) $ in our scheme.

The relevant mixing angles are  $\delta_{23}^{dL}$ and  $\delta_{23}^{dR}$
 for $B_s\rightarrow K^+K^-$ and 
$B_s\rightarrow K^0 \overline K^0$ decays. 
These mixing angles are  complex, and then we take
 \begin{equation}
  |\delta _{23}^{dR}|=|\delta _{23}^{dL}| \ ,
  \label{assumtion}
 \end{equation}
 for simplicity.
 On the other hand, the phases of $\delta_{23}^{dR}$ and $\delta_{23}^{dL}$ are   free parameters, which are 
 are constrained by experimental data.
 
 We comment on our assumption in Eq.(\ref{assumtion}).
 This one may be motivated from the $SO(10)$ GUT model with SUSY apart from  phases.
 In practice, this case of Eq.(\ref{assumtion}) give us the largest  SUSY contribution in our prediction
 because the  SUSY one is  symmetric for $\delta_{23}^{dR}$ and $\delta _{23}^{dL}$ in our framework.
 Therefore, our predicted region  of the CP violations is not changed even if this assumption is relaxed.

\section{Numerical Results}

We show  predicted numerical results of the CP violation in our framework.
Let us start with presenting  the SM prediction of the direct CP 
asymmetry of the $B^0\rightarrow K^+\pi^-$ process
\begin{align} 
A(\bar{B^0}\to K^-\pi^+)=
\frac{|\mathcal{\bar A}(\bar{B^0}\to K^-\pi^+)|^2-|\mathcal{A}({B^0}\to K^+\pi^-)|^2}
{|\mathcal{\bar A}(\bar{B^0}\to K^-\pi^+)|^2+|\mathcal{A}({B^0}\to K^+\pi^-)|^2} \ .
\end{align} 

 The predicted asymmetry  depends on $|V_{ub}|$ and $\gamma$ in the SM.
We show it  versus $|V_{ub}|$ in Figure 1(a), where
the recent measurements of $|V_{ub}|$ and $\gamma$ are taken as follows  \cite{Ciuchini-KEKFF2013}:
 \begin{align} 
|V_{ub}|=(3.82\pm 0.56)\times 10^{-3} \ , \qquad  \gamma=(70.8\pm 7.8)^\circ \ ,
\end{align} 
and other input parameters in our calculation  are summarized in Table 1.
\begin{table}[h]
\begin{center}
\begin{tabular}{|l|}
\hline
$\alpha_s(M_Z)=0.1184$ \cite{PDG}\\
$m_s(2{\rm GeV})=0.095$ GeV \cite{PDG}\\
$m_c(m_c)=1.275$ GeV \cite{PDG}\\
$m_b(m_b)=4.18$ GeV \cite{PDG}\\
$m_t(m_t)=160.0$ GeV $(\overline{{\text MS}})$ \cite{PDG}\\
$M_{B_s}=5.36677(24)$ GeV \cite{PDG}\\
$\Delta M_{B_s}=(116.942 \pm 0.1564)\times 10^{-13}$ GeV \cite{Aaij:2013mpa}\\
$f_{B_s} = (233\pm 10)$ MeV \cite{Ciuchini-KEKFF2013}\\
$f_\pi = (130.7\pm 0.4)$ MeV \cite{PDG}\\ %
$f_K = (156.1\pm1.1)$ MeV \cite{PDG}\\ 
$\lambda=0.2255(7)$ \cite{PDG}\\
$|V_{cb}|=(4.12\pm0.11)\times 10^{-2}$  \cite{Ciuchini-KEKFF2013}\\
\hline
\end{tabular}
\end{center}
\caption{Input parameters in our calculation.}
\label{tab:inputparameters}
\end{table}

As seen in Figure 1(a), the SM prediction  completely agrees with the
 observed value $-0.0082\pm 0.006$ \cite{Amhis:2012bh}.
 The predicted asymmetry is linear dependent  on $|V_{ub}|$.
 As far as $|V_{ub}|=(3.2-4.2)\times 10^{-3}$, our prediction  is successful.
  Our prediction is not sensitive to  $\gamma$ in the region of $\gamma=(70.8\pm 7.8)^\circ$ since
   $\sin \gamma$ is not so changed.
 More precise data of the asymmetry and $|V_{ub}|$ is crucial test of our  SM prediction with the QCD factorization.
 
 We also present the CP averaged branching ratio versus the form factor $F^{B^0\to K}(0)$
in Figure 1(b), in which the magnitude of the form factor is taken to be 
$F^{B^0\to K}(0)=0.26-0.42$ \cite{Hofer:2010ee}. 
The CP averaged branching ratio is also consistent with the observed one if $F^{B^0\to K}(0)=0.37-0.42$.
We omit  figures of the  $|V_{ub}|$ and  $\gamma$ dependences of the branching ratio
because it is insensitive to $|V_{ub}|$ and  $\gamma$.
\begin{figure}[h]
\includegraphics[width=7.8cm]{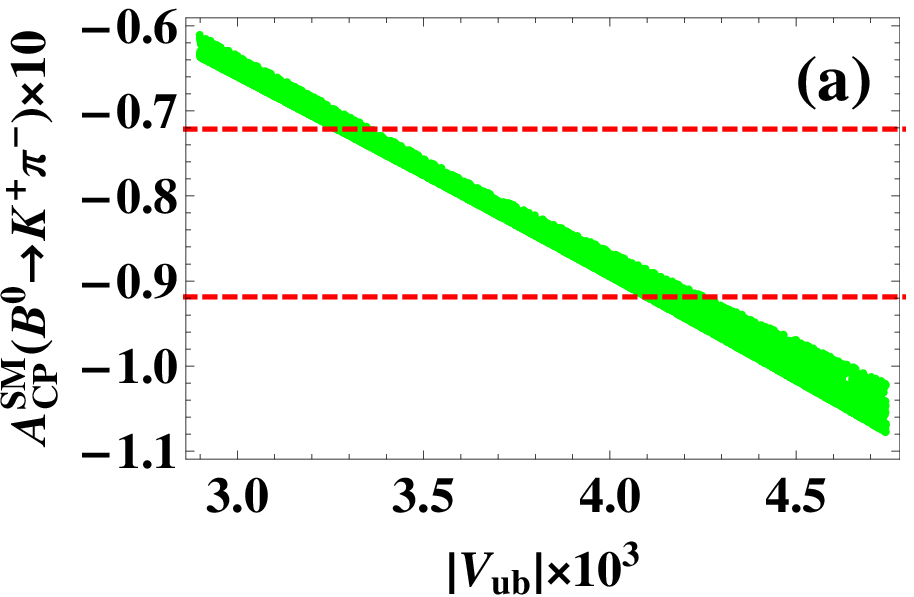}
\hspace{1cm}
\includegraphics[width=7.5cm]{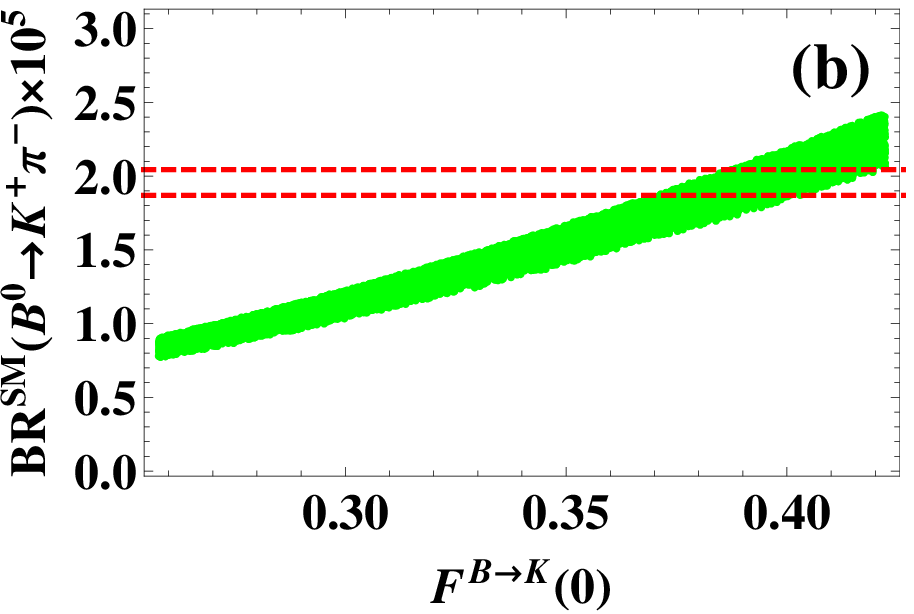}
\caption{Predictions of (a) the asymmetery versus  $|V_{ub}|$  and (b) the branching ratio versus the form factor $F^{B^0\to K}(0)$
  in  the  $B^0\rightarrow K^+\pi^-$ decay.
The inside between dashed red lines denotes the experimental allowed region at $90\%$C.L.} 
\end{figure}

 The agreement between the SM prediction and the experimental data  indicates that
 the SUSY contribution is constrained severely by the direct CP violation of $\bar{B^0}\to K^-\pi^+$.
We have searched the allowed parameter region of $\delta_{23}^{dL(dR)}$ by
 scattering  the magnitude of $\delta_{23}^{dL(dR)}$ 
 and these phases in the region of $0 \sim 0.1$ and $-\pi \sim  \pi$, respectively.
These parameters are constrained by
 the mass difference $\Delta M_{B_s}$, the CP violating phase  $\phi_s$ in
 $B_s\to J/\psi  \phi$ decay and  the branching ratio of the $b\to s\gamma$ decay.
 In addition to these data,  the asymmetry of $A(\bar{B^0}\to K^-\pi^+)$ constrains the magnitude of $\delta_{23}^{dL(dR)}$.
 We show the predicted asymmetry versus the magnitude of $\delta_{23}^{dL(dR)}$ in Figure 2,
 where its phase is taken in $-\pi\sim \pi$. It is  found that the SUSY contribution 
 becomes important in the region of  $|\delta_{23}^{dL(dR)}|\geq 0.01$.
 
 We also present  the predicted  branching ratio of the $b\to s\gamma$ decay versus the magnitude of $\delta_{23}^{dL(dR)}$
 in Figure 3.  The significant contribution of the SUSY effect is also seen in the region of  
$|\delta_{23}^{dL(dR)}|\geq 0.01$. 
 
\begin{figure}[h]
\begin{minipage}[]{0.45\linewidth}
\includegraphics[width=7.8cm]{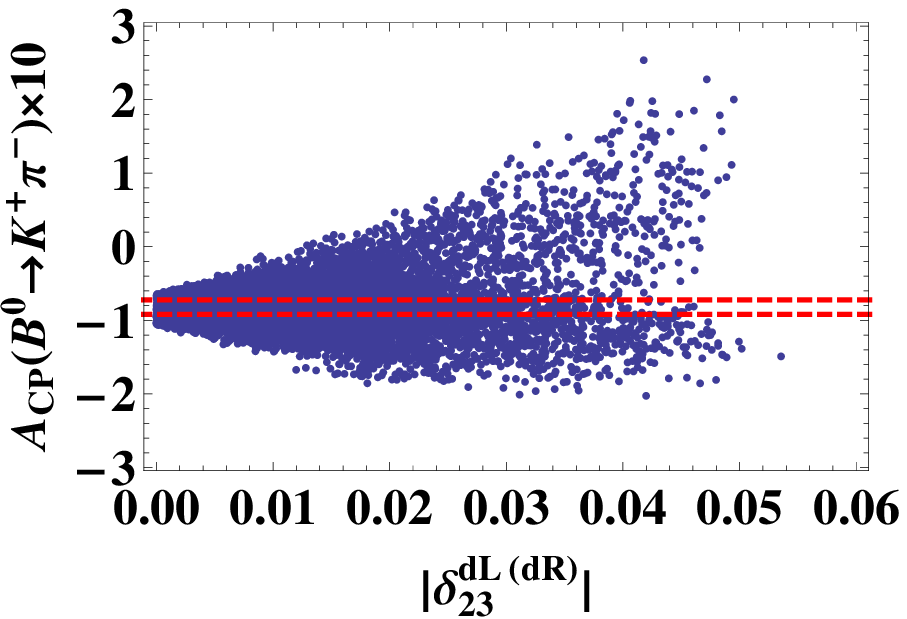}
\hspace{1cm}
\caption{The predicted $A(\bar{B^0}\to K^-\pi^+)$ versus  $|\delta_{23}^{dL(dR)}|$.
The inside between  dashed red lines denotes the experimental allowed region at $90\%$C.L.} 
\end{minipage}
\hspace{1cm}
\begin{minipage}[]{0.45\linewidth}
\includegraphics[width=7.5cm]{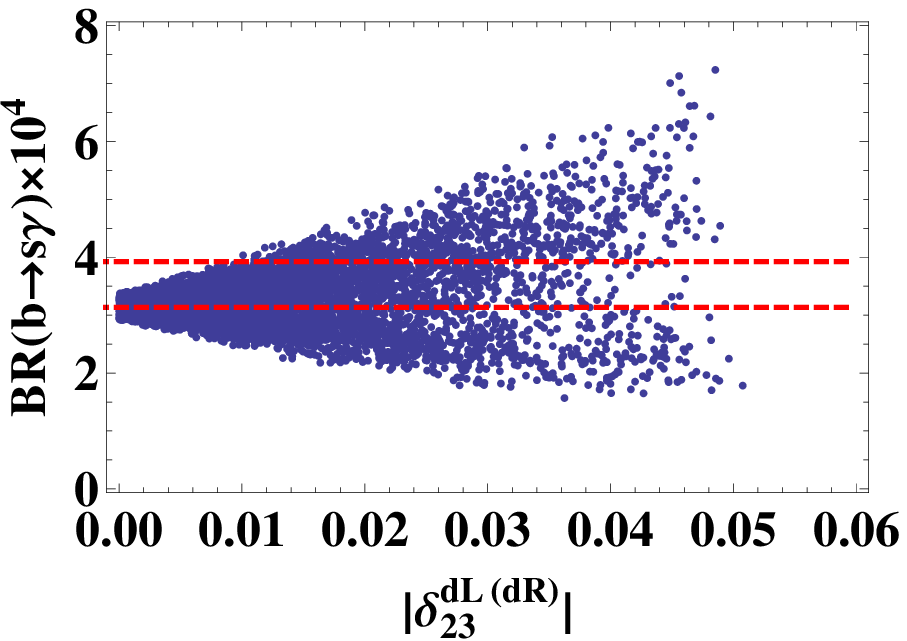}
\caption{ The predicted branching ratio of  $b\to s\gamma$ versus  $|\delta_{23}^{dL(dR)}|$.
The inside between dashed red lines denotes the experimental allowed region at $90\%$C.L.} 
\end{minipage}
\end{figure}
 
  Let us show the allowed region on the plane of $|\delta_{23}^{dL(R)}|$ and those phases,
   taking account of $\Delta M_{B_s}$, $\phi_s$ in
 $B_s\to J/\psi  \phi$ decay, the  branching ratio of $b\to s\gamma$, and the asymmetry  $A(\bar{B^0}\to K^-\pi^+)$.
 The input experimental data are taken at $90$ \% C.L.
 We present the allowed region of $|\delta_{23}^{dL(R)}|$ versus  $(\arg\delta_{23}^{dL}+\arg\delta_{23}^{dR})$ in
Figure 4(a), and versus  $(\arg\delta_{23}^{dL}-\arg\delta_{23}^{dR})$ in Figure 4(b) with
$|\delta_{23}^{dL}|=|\delta_{23}^{dR}|$, respectively.
 It is found that the squark flavor mixing is allowed in the region of    
 $|\delta_{23}^{dL}|\leq 0.02$ for all region of the  phase.
 If two phases $\arg\delta_{23}^{dL}$ and $\arg\delta_{23}^{dR}$ are tuned to suppress the imaginary part,
  $|\delta_{23}^{dL}|$ is allowed up to  $0.05$.

\begin{figure}[h]
\includegraphics[width=7.8cm]{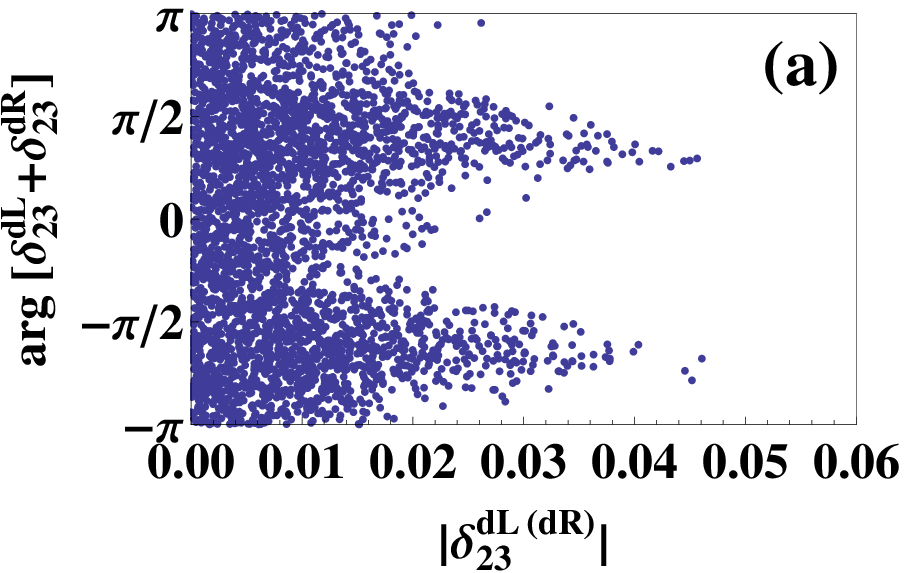}
\hspace{1cm}
\includegraphics[width=7.5cm]{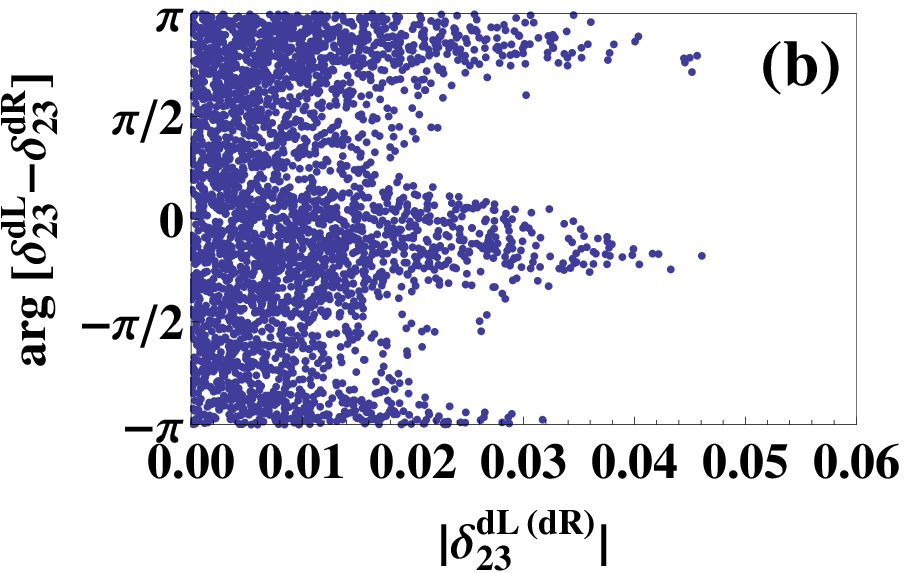}
\caption{The allowed region of  $|\delta_{23}^{dL(R)}|$
 versus (a) the sum of two phases and (b)  the difference of two phases.}
\end{figure}
\begin{figure}[h]
\includegraphics[width=7.8cm]{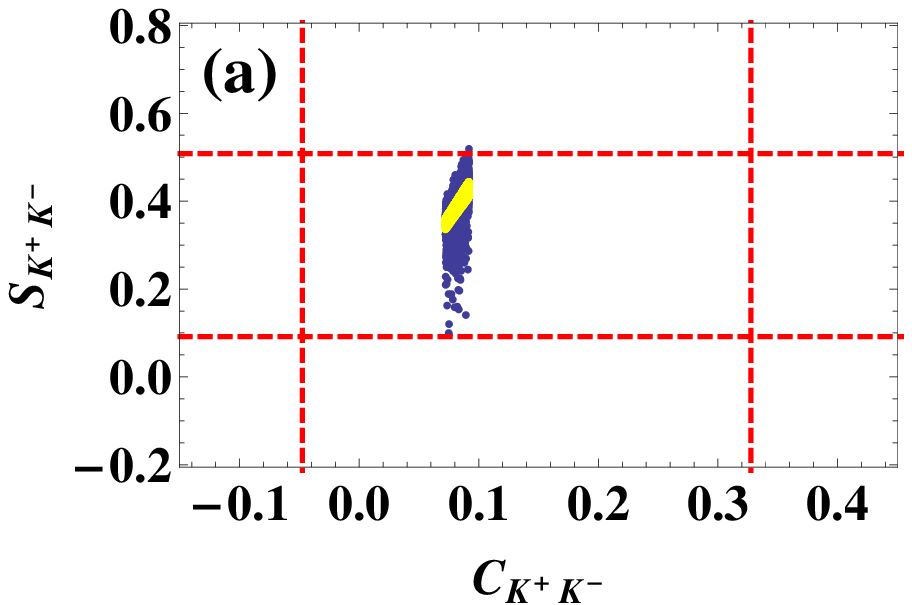}
\hspace{1cm}
\includegraphics[width=7.5cm]{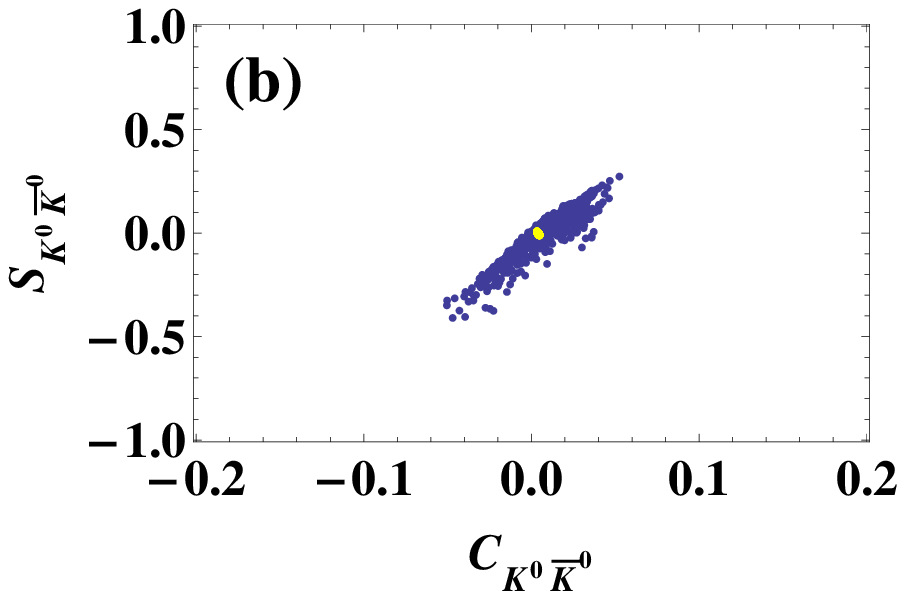}
\includegraphics[width=7.8cm]{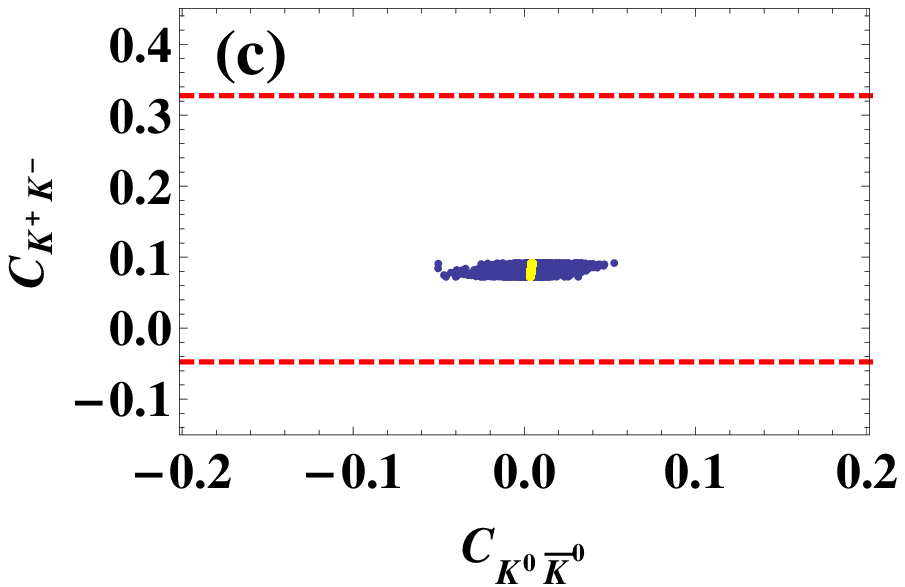}
\hspace{1cm}
\includegraphics[width=7.5cm]{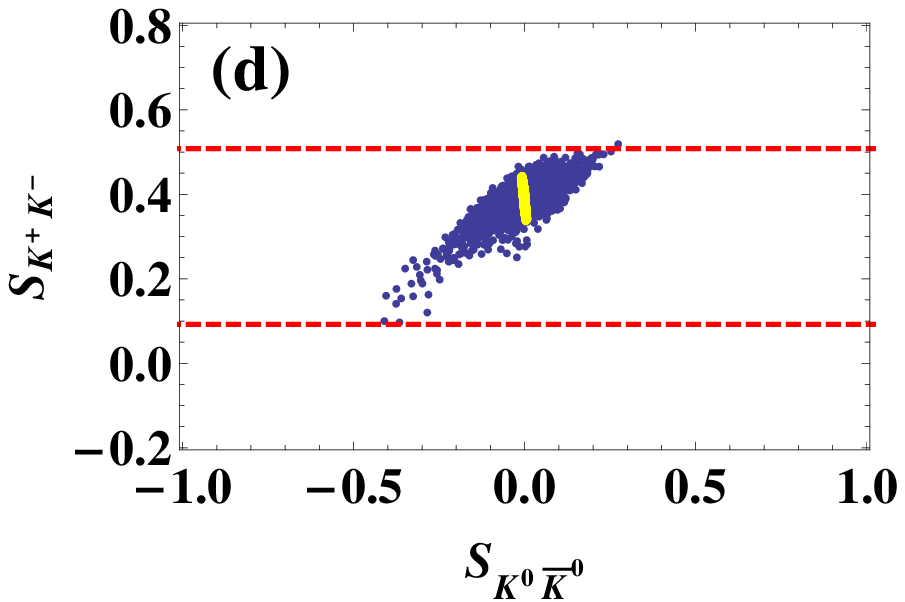}
\caption{The predicted CP violations  of  (a) $C_{K^+  K^-}-S_{K^+  K^-}$,  (b) $C_{K^0 \bar K^0}-S_{K^0 \bar K^0}$,
(c) $C_{K^0 \bar K^0}-C_{K^+  K^-}$, and (d) $S_{K^0 \bar K^0}-S_{K^+  K^-}$.
The inside between dashed red lines denotes the experimental allowed region at $90\%$C.L., and
yellow regions denote the SM predictions.} 
\end{figure}

 Now we can predict the CP violations of the  $B_s\rightarrow K^+K^-$ and
  $B_s\rightarrow K^0\overline K^0$ decays under the constraint of $\delta_{23}^{dL}$  of Figure 4.
  We show the predicted regions among  $C_{K^+K^-}$, $S_{K^+K^-}$, $C_{K^0 \bar K^0}$ and  $S_{K^0 \bar K^0}$
  in Figures 5(a)-5(d).  As seen in Figure 5(a),   the predicted region of $C_{K^+K^-}$ is strongly cut 
   by the constraint from the direct CP violation of $\bar{B^0}\to K^-\pi^+$.
  Therefore, the deviation from the SM prediction of $C_{K^+K^-}$ is not found.  On the other hand, 
 $S_{K^+K^-}$ is possibly deviated from the SM prediction considerably,
 that is expected to be in $0.1\sim 0.5$.
  The precise measurement of  $S_{K^+K^-}$  is important to search for the SUSY effect. 
   
   As seen in Figure 5(b),   the SM predictions of $C_{K^0 \bar K^0}$ and $S_{K^0 \bar K^0}$
   are very small since  we have 
 \begin{align} 
 \frac{\mathcal{\bar A}(\bar B_s\to K^0 \bar K^0)}{\mathcal{A}(B_s\to K^0 \bar K^0)}\simeq
 \frac{V_{tb}V^*_{ts}}{V^*_{tb}V_{ts}} \ ,  \qquad
 \frac{q}{p}\simeq \frac{V_{tb}^*V_{ts}}{V_{tb}V^*_{ts}}\ , \qquad \lambda_{ K^0 \bar K^0}\simeq 1 \ ,
\end{align} 
where the CKM matrix elements canceled out each other in $\lambda_{K^0 \bar K^0}$.
Since the SUSY contribution violates this cancellation,
 we expect the observation  of the CP violation for both $C_{K^0 \bar K^0}$ and $S_{K^0 \bar K^0}$
 in the $B_s\to K^0 \bar K^0$ decay.  These predicted  magnitudes are roughly  proportional to  each other
 in the region  $C_{K^0 \bar K^0}=-0.06\sim 0.06$ and $S_{K^0 \bar K^0}=-0.5\sim 0.3$.
 
 We show the correlations between $C_{K^0 \bar K^0}$ and $C_{K^+  K^-}$ in Figures 5(c),
  and between $S_{K^0 \bar K^0}$ and $S_{K^+  K^-}$ in Figures 5(d), respectively.
  While the predicted value of $C_{K^+  K^-}$ is restricted around $0.1$,
  $C_{K^0 \bar K^0}$ is expected in the region of $-0.06\sim 0.06$.
  On the other hand, $S_{K^0 \bar K^0}$ is roughly proportional to $S_{K^+  K^-}$, which
  gives  us   a crucial test for the SUSY contribution.

\section{Summary}
 In order to search for the gluino-squark mediated  flavor changing effect, we have studied the CP violations in the  $B_s\rightarrow K^+K^-$ and
  $B_s\rightarrow K^0\overline K^0$ processes, in which  the  $b\to s$ transition penguin amplitudes dominate the decays.
We have searched for the allowed  region of the flavor mixing $\delta_{23}^{dL}$,  by putting the experimental data 
 the mass difference $\Delta M_{B_s}$, the CP violating phase  $\phi_s$ in
 $B_s\to J/\psi  \phi$ decay and  the $b\to s\gamma$ branching ratio. 
In addition to these data, we  have taken into account  the constraint from  the asymmetry of $B^0\rightarrow K^+\pi^-$
because  the $B_s\rightarrow K^+K^-$ decay is related with the  $B^0\rightarrow K^+\pi^-$ decay 
by replacing the spectator $s$ with $d$.   We have obtained the constraint of  $|\delta_{23}^{dL}|\leq 0.05$.

 Under the constraint, we have  predicted the CP violations in the  $B_s\rightarrow K^+K^-$ and
  $B_s\rightarrow K^0\overline K^0$ decays. 
  The predicted region of the  CP violation $C_{K^+K^-}$ is strongly cut 
   by the constraint from the direct CP violation of $\bar{B^0}\to K^-\pi^+$,
   which is well agreement with the SM prediction with the QCD factorization calculation,
  Therefore, the deviation from the SM prediction of $C_{K^+K^-}$ is not expected.
  On the other hand, 
 $S_{K^+K^-}$ is possibly deviated from the SM prediction considerably, in the region of $0.1\sim 0.5$.
 Since the SM predictions of $C_{K^0 \bar K^0}$ and $S_{K^0 \bar K^0}$
   are tiny, the SUSY contribution is expected to be detectable 
in $C_{K^0 \bar K^0}$ and $S_{K^0 \bar K^0}$.  These expected magnitudes are
 in the region  $C_{K^0 \bar K^0}=-0.06\sim 0.06$ and $S_{K^0 \bar K^0}=-0.5\sim 0.3$.
 We expect  more precise data of the CP violations in  these decays, which  provide us a crucial test for the SUSY contribution.

\vspace{1 cm}
\noindent
{\bf Acknowledgment}

Y.S. is supported by JSPS Postdoctoral Fellowships 
for Research Abroad, No.20130600.
This work is also  supported by JSPS Grand-in-Aid for Scientific Research,
 21340055 and 24654062, 25-5222, respectively.
\appendix{}
\section*{Appendix}
\section{Loop integral in penguins}

The loop integrals in Eq.(\ref{ai}) are given as follows \cite{Muta:2000ti,Giri:2004af}:

\begin{align}
&F_{M_2}
=-12\ln\frac{\mu}{m_b}-18+f_{M_2}^{\rm I}+f_{M_2}^{\rm II}, \nonumber \\
&f_{M_2}^{\rm I}
=\int_0^1dx \ g(x)\phi_{M_2}(x), \qquad g(x)=3\frac{1-2x}{1-x}\ln x-3i\pi, \nonumber\\
&f_{M_2}^{\rm II}
=\frac{4 \pi^2}{N}\frac{f_{M_1}f_B}{f_+^{B\to M_1}(0)M_B^2}\int_0^1dz \frac{\phi_B(z)}{z}\int_0^1dx \frac{\phi_{M_1}(x)}{x}\int_0^1dy \frac{\phi_{M_2}(y)}{y}, \nonumber\\
&G_{M_{2},g}
=-\int_0^1dx \frac{2}{\bar x}\phi_{M_2}(x),\nonumber\\
&G_{M_{2}(s_q)}
=\frac{2}{3}-\frac{4}{3}\ln \frac{\mu}{m_b}+4\int_0^1dx\phi_{M_2}(x)\int_0^1du \ u\bar{u}\ln [s_q-u\bar u \bar x-i\epsilon],
\nonumber\\
&G_{M_{2},g}'
=-\int_0^1dx \frac{3}{2}\phi_{M_2}^0(x) \ =-\frac{3}{2},\nonumber\\
&G_{M_{2}}'(s_q)
=\frac{1}{3}-\ln \frac{\mu}{m_b}+3\int_0^1dx\phi_{M_2}^0(x)\int_0^1du \ u\bar u
\ln [s_q-u\bar u\bar x-i\epsilon],
\end{align} 
where $\bar{x}=1-x$ and $\bar{u}=1-u$.
The internal quark mass in the penguin diagrams enters as $s_f=m_f^2/m_b^2$. 
The functions $\phi(x)$ and $\phi^0(x)$ are meson's leading-twist distribution amplitude and twist-3 distribution amplitude, respectively.  For $\pi$ and $K$ mesons, we use well known form
\cite{Chernyak:1983ej,Braun:1989iv}:
\begin{align}
\phi_{\pi,K}(x)=6x (1-x) \ , \quad \phi_{\pi,K}^0(x)=1 \ .
\end{align}
For the $B$ meson, we use \cite{Keum:2000ph,Keum:2000wi,Yu:2013pua}
\begin{align}
\phi_{B}(x)=N_B x^2(1-x)^2 \exp\left [  -\frac{M_B^2 x^2}{2\omega_B^2}\right ] \ ,
\end{align}
where $\omega_B=0.4$GeV, and $0.5 GeV$ for the $B^0$ and $B_s$ mesons, respectively, and $N_B$ is the normalization constant
to make $\int^1_0 dx\phi_B(x)=1. $

\section{Squark contribution in $\Delta B=1$ process}

The Wilson coefficients for the gluino contribution 
in Eq.(\ref{hamiltonian}) are written as \cite{GotoNote}

\begin{align}
C_{7\gamma }^{\tilde g}(m_{\tilde g}) &= 
\frac{8}{3}\frac{\sqrt{2}\alpha _s\pi }{2G_FV_{tb}V_{tq}^*} \nonumber \\
&\times \Bigg [\frac{\big (\Gamma _{GL}^{(d)}\big )_{k3}^*}{m_{\tilde d_3}^2}
\left \{ \big (\Gamma _{GL}^{(d)}\big )_{33}\left (-\frac{1}{3}F_2(x_{\tilde g}^3)\right )
+\frac{m_{\tilde g}}{m_b}\big (\Gamma _{GR}^{(d)}\big )_{33}
\left (-\frac{1}{3}F_4(x_{\tilde g}^3)\right )\right \} \nonumber \\
&\hspace{3mm}+\frac{\big (\Gamma _{GL}^{(d)}\big )_{k6}^*}{m_{\tilde d_6}^2}
\left \{ \big (\Gamma _{GL}^{(d)}\big )_{36}\left (-\frac{1}{3}F_2(x_{\tilde g}^6)\right )
+\frac{m_{\tilde g}}{m_b}\big (\Gamma _{GR}^{(d)}\big )_{36}
\left (-\frac{1}{3}F_4(x_{\tilde g}^6)\right )\right \} \Bigg ],
\end{align}
\begin{align}
C_{8G}^{\tilde g}(m_{\tilde g}) &= \frac{8}{3}\frac{\sqrt{2}\alpha _s\pi }{2G_FV_{tb}V_{tq}^*}
\Bigg [\frac{\big (\Gamma _{GL}^{(d)}\big )_{k3}^*}{m_{\tilde d_3}^2}
\left \{ \big (\Gamma _{GL}^{(d)}\big )_{33}
\left (-\frac{9}{8}F_1(x_{\tilde g}^3)-\frac{1}{8}F_2(x_{\tilde g}^3)\right )\right .\nonumber \\
&\hspace{3.8cm}\left .+\frac{m_{\tilde g}}{m_b}\big (\Gamma _{GR}^{(d)}\big )_{33}
\left (-\frac{9}{8}F_3(x_{\tilde g}^3)-\frac{1}{8}F_4(x_{\tilde g}^3)\right )\right \} \nonumber \\
&\hspace{2.4cm}+\frac{\big (\Gamma _{GL}^{(d)}\big )_{k6}^*}{m_{\tilde d_6}^2}
\left \{ \big (\Gamma _{GL}^{(d)}\big )_{36}
\left (-\frac{9}{8}F_1(x_{\tilde g}^6)-\frac{1}{8}F_2(x_{\tilde g}^6)\right )\right .\nonumber \\
&\hspace{3.8cm}\left .+\frac{m_{\tilde g}}{m_b}\big (\Gamma _{GR}^{(d)}\big )_{36}
\left (-\frac{9}{8}F_3(x_{\tilde g}^6)-\frac{1}{8}F_4(x_{\tilde g}^6)\right )\right \} \Bigg ],
\end{align}
where $k=2,1$ correspond to $b\to q~(q=s,d)$ transitions, respectively. 
The loop functions $F_i(x_{\tilde g}^I)$ are given as 
\begin{align}
F_1(x_{\tilde g}^I)&=\frac{x_{\tilde g}^I\log x_{\tilde g}^I}{2(x_{\tilde g}^I-1)^4}
+\frac{(x_{\tilde g}^I)^2-5x_{\tilde g}^I-2}{12(x_{\tilde g}^I-1)^3}~,\nonumber \\
F_2(x_{\tilde g}^I)&=-\frac{(x_{\tilde g}^I)^2\log x_{\tilde g}^I}{2(x_{\tilde g}^I-1)^4}
+\frac{2(x_{\tilde g}^I)^2+5x_{\tilde g}^I-1}{12(x_{\tilde g}^I-1)^3}~,\nonumber \\
F_3(x_{\tilde g}^I)&=\frac{\log x_{\tilde g}^I}{(x_{\tilde g}^I-1)^3}
+\frac{x_{\tilde g}^I-3}{2(x_{\tilde g}^I-1)^2}~,\nonumber \\
F_4(x_{\tilde g}^I)&=-\frac{x_{\tilde g}^I\log x_{\tilde g}^I}{(x_{\tilde g}^I-1)^3}+
\frac{x_{\tilde g}^I+1}{2(x_{\tilde g}^I-1)^2}=\frac{1}{2}g_{2[1]}(x_{\tilde g}^I,x_{\tilde g}^I)~,
\end{align}
with $x_{\tilde g}^I=m_{\tilde g}^2/m_{\tilde d_I}^2~(I=3,6)$.
The NLO of these Wilson coefficients are   omitted.
We also omit other  Wilson coefficients which  are
 the NLO contributions to our numerical calculations.
The Wilson coefficients $\widetilde C_i^{\tilde g}(m_{\tilde g})$'s are 
obtained by replacing $L(R)$ with $R(L)$ in $C_i^{\tilde g}(m_{\tilde g})$'s.

The Wilson coefficients of $C_{7\gamma}^{\tilde g}(m_b)$ and 
$C_{8G}^{\tilde g}(m_b)$ at the $m_b$ scale are given at the leading order of QCD as follows~\cite{Buchalla:1995vs}: 
\begin{equation}
\begin{split}
C_{7\gamma}^{\tilde g}(m_b)
&= \zeta C_{7\gamma}^{\tilde g}(m_{\tilde g})
+\frac{8}{3}(\eta-\zeta) C_{8G}^{\tilde g}(m_{\tilde g}), \cr
C_{8G}^{\tilde g}(m_b)
&=\eta C_{8G}^{\tilde g}(m_{\tilde g}),
\end{split}
\end{equation}
where 
\begin{equation}
\zeta=\left ( 
 \frac{\alpha_s(m_{\tilde g})}{\alpha_s(m_t)} \right )^{\frac{16}{21}}
 \left ( 
 \frac{\alpha_s(m_t)}{\alpha_s(m_b)} \right )^{\frac{16}{23}} \ , \qquad
 \eta=\left ( 
 \frac{\alpha_s(m_{\tilde g})}{\alpha_s(m_t)} \right )^{\frac{14}{21}}
 \left ( 
 \frac{\alpha_s(m_t)}{\alpha_s(m_b)} \right )^{\frac{14}{23}} \ .
 \end{equation}

\section{Squark contribution in $\Delta B=2$ process}

The $\Delta B=2$ effective Lagrangian from the gluino-sbottom-quark interaction  is given as
\begin{align}
\mathcal{L}_{\text{eff}}^{\Delta F=2}=&-\frac{1}{2}\left [C_{VLL}O_{VLL}+C_{VRR}O_{VRR}\right ] \nonumber \\
&-\frac{1}{2}\sum _{i=1}^2
\left [C_{SLL}^{(i)}O_{SLL}^{(i)}+C_{SRR}^{(i)}O_{SRR}^{(i)}+C_{SLR}^{(i)}O_{SLR}^{(i)}\right ],
\label{Lagrangian-DeltaF=2}
\end{align}
then, the $P^0$-$\bar P^0$ mixing, $M_{12}$, is written as 
\begin{equation}
M_{12}=-\frac{1}{2m_P}\langle P^0|\mathcal{L}_{\text{eff}}^{\Delta F=2}|\bar P^0\rangle \ .
\end{equation}
The hadronic matrix elements are given in terms of the non-perturbative
parameters  $B_i$ as: 
\begin{align}
\langle P^0|\mathcal{O}_{VLL}|\bar P^0\rangle &=\frac{2}{3}m_P^2f_P^2B_1, \quad 
\langle P^0|\mathcal{O}_{VRR}|\bar P^0\rangle =\langle P^0|\mathcal{O}_{VLL}|\bar P^0\rangle ,\nonumber \\
\langle P^0|\mathcal{O}_{SLL}^{(1)}|\bar P^0\rangle &=-\frac{5}{12}m_P^2f_P^2R_PB_2, \quad 
\langle P^0|\mathcal{O}_{SRR}^{(1)}|\bar P^0\rangle =\langle P^0|\mathcal{O}_{SLL}^{(1)}|\bar P^0\rangle ,\nonumber \\
\langle P^0|\mathcal{O}_{SLL}^{(2)}|\bar P^0\rangle &=\frac{1}{12}m_P^2f_P^2R_PB_3, \quad 
\langle P^0|\mathcal{O}_{SRR}^{(2)}|\bar P^0\rangle =\langle P^0|\mathcal{O}_{SLL}^{(2)}|\bar P^0\rangle ,\nonumber \\
\langle P^0|\mathcal{O}_{SLR}^{(1)}|\bar P^0\rangle &=\frac{1}{2}m_P^2f_P^2R_PB_4, \quad 
\langle P^0|\mathcal{O}_{SLR}^{(2)}|\bar P^0\rangle =\frac{1}{6}m_P^2f_P^2R_PB_5,
\end{align}
where 
\begin{equation}
R_P=\left (\frac{m_P}{m_Q+m_q}\right )^2,
\end{equation}
with $(P,Q,q)=(B_d,b,d),~(B_s,b,s)$.

The Wilson coefficients for the gluino contribution in Eq.~(\ref{Lagrangian-DeltaF=2}) are written as \cite{GotoNote}

\begin{align}
C_{VLL}(m_{\tilde g})&=\frac{\alpha _s^2}{m_{\tilde g}^2}\sum _{I,J=1}^6
(\lambda _{GLL}^{(d)})_I^{ij}(\lambda _{GLL}^{(d)})_J^{ij}
\left [\frac{11}{18}g_{2[1]}(x_I^{\tilde g},x_J^{\tilde g})
+\frac{2}{9}g_{1[1]}(x_I^{\tilde g},x_J^{\tilde g})\right ],\nonumber \\
C_{VRR}(m_{\tilde g})&=C_{VLL}(m_{\tilde g})(L\leftrightarrow R),\nonumber \\
C_{SRR}^{(1)}(m_{\tilde g})&=\frac{\alpha _s^2}{m_{\tilde g}^2}\sum _{I,J=1}^6
(\lambda _{GLR}^{(d)})_I^{ij}(\lambda _{GLR}^{(d)})_J^{ij}
\frac{17}{9}g_{1[1]}(x_I^{\tilde g},x_J^{\tilde g}),\nonumber \\
C_{SLL}^{(1)}(m_{\tilde g})&=C_{SRR}^{(1)}(m_{\tilde g})(L\leftrightarrow R),\nonumber \\
C_{SRR}^{(2)}(m_{\tilde g})&=\frac{\alpha _s^2}{m_{\tilde g}^2}\sum _{I,J=1}^6
(\lambda _{GLR}^{(d)})_I^{ij}(\lambda _{GLR}^{(d)})_J^{ij}
\left (-\frac{1}{3}\right )g_{1[1]}(x_I^{\tilde g},x_J^{\tilde g}),\nonumber \\
C_{SLL}^{(2)}(m_{\tilde g})&=C_{SRR}^{(2)}(m_{\tilde g})(L\leftrightarrow R),\nonumber \\
C_{SLR}^{(1)}(m_{\tilde g})&=\frac{\alpha _s^2}{m_{\tilde g}^2}\sum _{I,J=1}^6
\Bigg \{ (\lambda _{GLR}^{(d)})_I^{ij}(\lambda _{GRL}^{(d)})_J^{ij}
\left (-\frac{11}{9}\right )g_{2[1]}(x_I^{\tilde g},x_J^{\tilde g}) \nonumber \\
&\hspace{2cm}+(\lambda _{GLL}^{(d)})_I^{ij}(\lambda _{GRR}^{(d)})_J^{ij}
\left [\frac{14}{3}g_{1[1]}(x_I^{\tilde g},x_J^{\tilde g})
-\frac{2}{3}g_{2[1]}(x_I^{\tilde g},x_J^{\tilde g})\right ]\Bigg \} ,\nonumber \\
C_{SLR}^{(2)}(m_{\tilde g})&=\frac{\alpha _s^2}{m_{\tilde g}^2}\sum _{I,J=1}^6
\Bigg \{ (\lambda _{GLR}^{(d)})_I^{ij}(\lambda _{GRL}^{(d)})_J^{ij}
\left (-\frac{5}{3}\right )g_{2[1]}(x_I^{\tilde g},x_J^{\tilde g}) \nonumber \\
&\hspace{2cm}+(\lambda _{GLL}^{(d)})_I^{ij}(\lambda _{GRR}^{(d)})_J^{ij}
\left [\frac{2}{9}g_{1[1]}(x_I^{\tilde g},x_J^{\tilde g})
+\frac{10}{9}g_{2[1]}(x_I^{\tilde g},x_J^{\tilde g})\right ]\Bigg \} ,
\end{align}
where
\begin{align}
(\lambda _{GLL}^{(d)})_K^{ij}&=(\Gamma _{GL}^{(d)\dagger })_i^K(\Gamma _{GL}^{(d)})_K^j~,\quad 
(\lambda _{GRR}^{(d)})_K^{ij}=(\Gamma _{GR}^{(d)\dagger })_i^K(\Gamma _{GR}^{(d)})_K^j~,\nonumber \\
(\lambda _{GLR}^{(d)})_K^{ij}&=(\Gamma _{GL}^{(d)\dagger })_i^K(\Gamma _{GR}^{(d)})_K^j~,\quad 
(\lambda _{GRL}^{(d)})_K^{ij}=(\Gamma _{GR}^{(d)\dagger })_i^K(\Gamma _{GL}^{(d)})_K^j~\ .
\end{align}
Here we take $(i,j)=(1,3),~(2,3)$ which correspond to the $B^0$ and $B_s$ mesons, respectively. 
The loop functions are given as follows:
\begin{itemize}
\item If $x_I^{\tilde g}\not =x_J^{\tilde g}$ ($x_{I,J}^{\tilde g}=m_{\tilde d_{I,J}}^2/m_{\tilde g}^2$),
\begin{align}
g_{1[1]}(x_I^{\tilde g},x_J^{\tilde g})&=\frac{1}{x_I^{\tilde g}-x_J^{\tilde g}}
\left (\frac{x_I^{\tilde g}\log x_I^{\tilde g}}{(x_I^{\tilde g}-1)^2}
-\frac{1}{x_I^{\tilde g}-1}-\frac{x_J^{\tilde g}\log x_J^{\tilde g}}{(x_J^{\tilde g}-1)^2}
+\frac{1}{x_J^{\tilde g}-1}\right ),\nonumber \\
g_{2[1]}(x_I^{\tilde g},x_J^{\tilde g})&=\frac{1}{x_I^{\tilde g}-x_J^{\tilde g}}
\left (\frac{(x_I^{\tilde g})^2\log x_I^{\tilde g}}{(x_I^{\tilde g}-1)^2}
-\frac{1}{x_I^{\tilde g}-1}-\frac{(x_J^{\tilde g})^2\log x_J^{\tilde g}}{(x_J^{\tilde g}-1)^2}
+\frac{1}{x_J^{\tilde g}-1}\right ).
\end{align}
\item If $x_I^{\tilde g}=x_J^{\tilde g}$\ ,
\begin{align}
g_{1[1]}(x_I^{\tilde g},x_I^{\tilde g})&=
-\frac{(x_I^{\tilde g}+1)\log x_I^{\tilde g}}{(x_I^{\tilde g}-1)^3}+\frac{2}{(x_I^{\tilde g}-1)^2}~,\nonumber \\
g_{2[1]}(x_I^{\tilde g},x_I^{\tilde g})&=
-\frac{2x_I^{\tilde g}\log x_I^{\tilde g}}{(x_I^{\tilde g}-1)^3}+\frac{x_I^{\tilde g}+1}{(x_I^{\tilde g}-1)^2}~.
\end{align}
\end{itemize}
In this paper, we take $(I,J)=(3,3),~(3,6),~(6,3),~(6,6)$, because we assume the split-family. 
The effective Wilson coefficients are given at the leading order of QCD as follows:
\begin{align}
C_{VLL}(m_b)=&\eta _{VLL}^{B}C_{VLL}(m_{\tilde g})\ ,\quad 
C_{VRR}(m_b)=\eta _{VRR}^{B}C_{VLL}(m_{\tilde g})\ ,\nonumber \\
\begin{pmatrix}
C_{SLL}^{(1)}(m_b) \\
C_{SLL}^{(2)}(m_b)
\end{pmatrix}&=
\begin{pmatrix}
C_{SLL}^{(1)}(m_{\tilde g}) \\
C_{SLL}^{(2)}(m_{\tilde g})
\end{pmatrix}X_{LL}^{-1}\eta _{LL}^{B}X_{LL} \ ,\nonumber \\
\begin{pmatrix}
C_{SRR}^{(1)}(m_b) \\
C_{SRR}^{(2)}(m_b)
\end{pmatrix}&=
\begin{pmatrix}
C_{SRR}^{(1)}(m_{\tilde g}) \\
C_{SRR}^{(2)}(m_{\tilde g})
\end{pmatrix}X_{RR}^{-1}\eta _{RR}^{B}X_{RR} \ ,\nonumber \\
\begin{pmatrix}
C_{SLR}^{(1)}(m_b) \\
C_{SLR}^{(2)}(m_b)
\end{pmatrix}&=
\begin{pmatrix}
C_{SLR}^{(1)}(m_{\tilde g}) \\
C_{SLR}^{(2)}(m_{\tilde g})
\end{pmatrix}X_{LR}^{-1}\eta _{LR}^{B}X_{LR}\ ,
\end{align}
where 
\begin{align}
&\eta _{VLL}^B=\eta _{VRR}^B=\left (\frac{\alpha _s(m_{\tilde g})}{\alpha _s(m_t)}\right )^{\frac{6}{21}}
\left (\frac{\alpha _s(m_t)}{\alpha _s(m_b)}\right )^{\frac{6}{23}}\ ,\nonumber \\
&\eta _{LL}^B=\eta _{RR}^B=
S_{LL}
\begin{pmatrix}
\eta _{b\tilde g}^{d_{LL}^1} & 0 \\
0 & \eta _{b\tilde g}^{d_{LL}^2}
\end{pmatrix}
S_{LL}^{-1}\ ,\qquad 
\eta _{LR}^B=S_{LR}
\begin{pmatrix}
\eta _{b\tilde g}^{d_{LR}^1} & 0 \\
0 & \eta _{b\tilde g}^{d_{LR}^2}
\end{pmatrix}
S_{LR}^{-1}\ ,\nonumber \\
&\eta _{b\tilde g}=\left (\frac{\alpha _s(m_{\tilde g})}{\alpha _s(m_t)}\right )^{\frac{1}{14}}
\left (\frac{\alpha _s(m_t)}{\alpha _s(m_b)}\right )^{\frac{3}{46}}\ ,\nonumber \\
\end{align}
\begin{align}
&d_{LL}^1=\frac{2}{3}(1-\sqrt{241}),\qquad d_{LL}^2=\frac{2}{3}(1+\sqrt{241})\ ,\qquad 
d_{LR}^1=-16,\qquad d_{LR}^2=2,\nonumber \\
&S_{LL}=
\begin{pmatrix}
\frac{16+\sqrt{241}}{60} & \frac{16-\sqrt{241}}{60} \\
1 & 1
\end{pmatrix}\ ,\quad 
S_{LR}=
\begin{pmatrix}
-2 & 1 \\
3 & 0
\end{pmatrix},\nonumber \\
&X_{LL}=X_{RR}=
\begin{pmatrix}
1 & 0 \\
4 & 8
\end{pmatrix},\qquad 
X_{LR}=
\begin{pmatrix}
0 & -2 \\
1 & 0
\end{pmatrix}\ .\nonumber \\
\end{align}


For the parameters $B_i^{(s)}(i=2-5)$ of the $B_s$ meson,  we use values in  \cite{Becirevic:2001xt}
as follows:
\begin{eqnarray}
&&B_2^{(B_s)} (m_b)=0.80(1)(4), \qquad
B_3^{(B_s)} (m_b)=0.93(3)(8), \nonumber\\
&&B_4^{(B_s)} (m_b)=1.16(2)(^{+5}_{-7}), \qquad
B_5^{(B_s)} (m_b)=1.75(3)(^{+21}_{-6})\ .
\end{eqnarray}
On the other hand, we use the most updated value for 
 $\hat B_1^{(s)} $ as \cite{Ciuchini-KEKFF2013,Flynn-KEKFF2013}
\begin{equation}
\hat B_1^{(B_s)}  = 1.33\pm 0.06\  .
\end {equation}


\end{document}